\documentclass[a4paper,11pt]{article}
\usepackage{pos}
\usepackage{graphicx}
\usepackage{graphics}
\usepackage{dcolumn}
\usepackage{multirow}

\usepackage{amsfonts}
\usepackage{amsmath}
\usepackage{bm}
\usepackage{mathtools}
\DeclareMathAlphabet{\mathcal}{OMS}{cmsy}{m}{n}

\newcommand{\Nf}{{N_f}}
\newcommand{\psibar}{{\overline{\psi}}}

\newcommand{\cZ}{{\cal Z}}
\newcommand{\MSbar}{{\overline{\text{MS}}}}
\newcommand{\corr}[1]{{\langle#1\rangle}}
\newcommand{\corrno}[1]{{\langle#1\rangle}} 
\newcommand{\bs}[1]{{\boldsymbol{#1}}}

\newcommand{\vxi}{{\bs \xi}}

\newcommand{\mt}{{\widetilde m}}

\newcommand{\mtht}[1]{{\texorpdfstring{#1}{}}}

\newcommand{\static}{{\frac{\Delta f_\vxi^\infty}{\Delta\xi_k}}}
\newcommand{\dynamic}{{\frac{\Delta (f_\vxi-f_\vxi^\infty)}{\Delta\xi_k}}}

\def\nab#1{{\nabla_{#1}}}
\def\nabstar#1{{\nabla\kern0.5pt\smash{\raise 4.5pt\hbox{$\ast$}}
               \kern-5.5pt_{#1}}}

\usepackage{xcolor}

\title{Non-perturbative determination of the QCD Equation of State up to the electroweak scale}

\author*[a]{Michele Pepe}

\affiliation[a]{INFN, Sezione di Milano-Bicocca, Piazza della Scienza 3, I-20126 Milano, Italy}

\emailAdd{michele.pepe@mib.infn.it}

\abstract{
The QCD Equation of State with $N_f=3$ massless quark flavours is determined non-perturbatively over a broad range of
temperatures, extending from the electroweak scale down to 3 GeV, and smoothly connecting to the low-temperature regime.
The comparison with perturbative predictions shows that, even at temperatures approaching the electroweak scale, the Equation of State can
be accurately described only by adding terms beyond the known perturbative series, including non-perturbative contributions. 
The strategy that allows this investigation in the previously unexplored high-temperature regime combines shifted boundary conditions
with a determination of the lines of constant physics based on the running of a non-perturbatively defined renormalized coupling. This
methodology is general and can be applied to QCD with four or five massive quark flavours.}

\FullConference{The 42nd International Symposium on Lattice Field Theory (LATTICE2025)\\
2-8 November 2025\\
Tata Institute of Fundamental Research, Mumbai, India\\}

\begin{document}
\maketitle
\section{Introduction}

The Equation of State (EoS) of Quantum Chromodynamics (QCD) characterizes the thermodynamic properties of strongly interacting matter in thermal
equilibrium. Its determination is a cornerstone of modern physics, bridging the gap between the microscopic dynamics of quarks and gluons
and the macroscopic cosmological evolution. While the EoS is a crucial input for hydrodynamic models of the quark-gluon plasma
generated in heavy-ion collisions at the hadronic scale ($T\sim$ 150 MeV), its relevance extends far beyond this regime.
In the early Universe, as the temperature decreased from the electroweak scale ($T \sim 100$~GeV) toward the QCD crossover, the QCD EoS
governed the expansion rate and the evolution of the effective number of relativistic degrees of freedom, with direct implications for the
spectrum of primordial gravitational waves~\cite{Saikawa:2018rcs}.  

Despite its critical importance, first-principles lattice determinations of the QCD EoS have so far been limited up to temperatures of about
1 GeV for theories with $N_f=2+1$~\cite{Borsanyi:2013bia,HotQCD:2014kol,Bazavov:2017dsy} or $N_f=2+1+1$~\cite{Borsanyi:2016ksw}. At higher
temperatures, our understanding has relied almost exclusively on perturbative expansions in the strong coupling constant $g$. However, the
convergence of the perturbative series is notoriously poor and at the order $g^6$ for the pressure~\cite{Kajantie:2002wa}, the expansion
becomes sensitive to ultrasoft modes that are inherently non-perturbative. Previous studies in SU(3) Yang-Mills theory~\cite{Giusti:2016iqr}
and investigations of QCD screening masses~\cite{DallaBrida:2021ddx,Giusti:2024ohu} indicate that perturbative predictions remain
significantly far from the full result even at temperatures approaching the electroweak scale. This suggests that the smooth approach to the
Stefan-Boltzmann limit may hide complex dynamics with non-perturbative contributions that can be resolved only through first-principles calculations.  

In these proceedings, we present an overview of the strategy and the results of a fully non-perturbative computation of the QCD EoS across
an unprecedented range of temperatures, from 3 GeV up to 165 GeV~\cite{Bresciani:2025vxw,Bresciani:2025mcu}. We focus on the theory with
$N_f=3$ massless quarks, a choice justified by the negligible impact of the light quark masses at these energy scales. This progress relies
on two key issues. First, the scale-setting problem is addressed through non-perturbative renormalization techniques in finite-volume schemes combined
with step-scaling methods, which allowed us to define lines of constant physics by following the running of a renormalized coupling up to very
high energies. Second, QCD is formulated in a moving reference frame through shifted boundary conditions in the Euclidean time
direction~\cite{Giusti:2010bb,Giusti:2011kt,Giusti:2012yj}. This allows for a direct determination of the entropy density and eliminates the
need for large ultraviolet subtractions of vacuum contributions. 

The methodology discussed here is general and provides a framework that can be directly applied to QCD with four or five massive quark flavors.

\section{Theoretical Framework}

We consider Quantum Chromodynamics with $N_f=3$ degenerate quark flavours regularized on a Euclidean lattice of size $L_0/a\times (L/a)^3$,
where $a$ is the lattice spacing. The action of the lattice theory $S_{QCD} = S_G + S_F$ is the sum of a pure gauge part, $S_G$, and a
fermionic one, $S_F$. 
For the gauge sector we consider the Wilson plaquette action~\cite{Wilson:1974sk},
\begin{equation}
    S_G = \frac{6}{g_0^2}\sum_x\sum_{\mu<\nu}\left[ 1 - \frac{1}{3}{\rm Re}\,{\rm tr}\left\{U_{\mu\nu}(x)\right\}\right]\,,
    \label{eq:SG}
\end{equation}
with $g_0$ being the bare coupling and $U_{\mu\nu}$ the plaquette field
\begin{equation}
  U_{\mu\nu}(x)=U_\mu(x)\, U_\nu(x+ a \hat{\mu})\, U_\mu(x+ a \hat{\nu})^\dag\, U_\nu(x)^\dag\,,
  \label{eq:plaquette}
\end{equation}
defined starting from the link field $U_\mu(x)\in$ SU(3). The fermionic action is
\begin{equation}
    S_F=a^4\sum_x \psibar(x) (D+M_0)\psi(x) \,,
    \label{eq:SF}
\end{equation}
where the fermionic (antifermionic) fields $\psi\,(\psibar)$ are triplets in color and flavour space and $M_0=m_0 1\!\!1$ is the bare mass
matrix of $N_f=3$ degenerate quarks. The lattice Dirac operator is defined as
\begin{equation}
    D=D_{\rm w} + a D_{\rm sw}\,,
    \label{eq:Dirac}
\end{equation}
where $D_{\rm w}$ is the massless Wilson-Dirac operator~\cite{Wilson:1975hf}
\begin{equation}
  D_{\rm w} = \frac{1}{2}\big\{\gamma_\mu(\nabstar\mu+\nab\mu)-a\nabstar\mu \nab\mu\big\}\,,
\end{equation}
and $D_{\rm sw}$ is the Sheikholeslami-Wohlert improvement term~\cite{Sheikholeslami:1985ij}
\begin{equation}
    D_{\rm sw}\psi(x) = c_{\rm sw}(g_0) \frac{1}{4}
    \sigma_{\mu\nu} \widehat F_{\mu\nu}(x)\psi(x) \,,
    \label{eq:DiracSW}
\end{equation}
with $\sigma_{\mu\nu}=\frac{i}{2}[\gamma_\mu,\gamma_\nu]$. The coefficient $c_{\mathrm{sw}}(g_0)$ is tuned
non-perturbatively~\cite{Yamada:2004ja} to remove leading O$(a)$ discretization effects~\cite{Sheikholeslami:1985ij,Luscher:1996sc}. 
The covariant lattice derivatives act on the quark fields as 
\begin{align}
    a \nab\mu \psi(x) & =  U_\mu(x)\psi(x+ a \hat{\mu})-\psi(x)\; ,\nonumber \\[0.25cm]
    a \nabstar\mu \psi(x) & = \psi(x) - U_\mu(x- a \hat{\mu})^\dag\psi(x - a \hat{\mu})\,,
\end{align} 
while the lattice field strength tensor $\widehat F_{\mu\nu}(x)$ is defined by the clover discretization
\begin{equation}
    \widehat  F_{\mu\nu}(x) = \frac{i}{8a^2}\big\{Q_{\mu\nu}(x)-Q_{\nu\mu}(x)\big\}\,,
    \label{eq:CloverFmunu}
\end{equation}
with
\begin{eqnarray}
  Q_{\mu\nu}(x) &=& U_\mu(x)U_\nu(x+a \hat{\mu})U_\mu(x+a \hat{\nu})^\dag U_\nu(x)^\dag\nonumber \\
  &+& U_\nu(x)U_\mu(x-a \hat{\mu}+a \hat{\nu})^\dag U_\nu(x-a \hat{\mu})^\dag U_\mu(x-a \hat{\mu})\nonumber \\
  &+& U_\mu(x-a \hat{\mu})^\dag U_\nu(x-a \hat{\mu}-a \hat{\nu})^\dag U_\mu(x-a \hat{\mu}-a \hat{\nu})U_\nu(x-a \hat{\nu})\nonumber \\
  &+& U_\nu(x-a \hat{\nu})^\dag U_\mu(x-a \hat{\nu})U_\nu(x+a \hat{\mu}-a \hat{\nu})U_\mu(x)^\dag\, .
\end{eqnarray}
Periodic boundary conditions are imposed in the spatial directions for both the gauge and the fermion fields. In the temporal direction, we
instead implement shifted boundary conditions~\cite{Giusti:2010bb,Giusti:2011kt,Giusti:2012yj}
\begin{equation}
\begin{aligned}
    U_\mu(x_0+L_0, \bs{x}) &= U_\mu(x_0, \bs{x}-L_0\vxi )\,, \\
    \psi(x_0+L_0, \bs{x})  &= -\psi(x_0, \bs{x}-L_0\vxi )\,, \\
    \psibar(x_0+L_0, \bs{x}) &= -\psibar(x_0, \bs{x}-L_0\vxi )\,.
\end{aligned}
\label{eq:shBCs}
\end{equation}
The spatial vector $\vxi$ represents a constant shift that generalizes the standard thermal boundary conditions that are recovered for
$\vxi=0$. A non-vanishing shift corresponds to formulating the theory in a moving reference frame and plays a central role in the
computation of thermodynamic observables. The partition function and thermal expectation values are defined by 
\begin{equation}
  \cZ_\vxi = \int DUD\psibar D\psi\, e^{-S_{QCD}}\,, \quad \langle O \rangle_{\bm{\xi}} =\dfrac{1}{\cZ_\vxi}
  \int DUD\psibar D\psi\, O(U,\psibar,\psi) \, e^{-S_{QCD}}.
\end{equation}

\subsection{Lines of Constant Physics}
A central challenge in high-temperature lattice simulations is the precise determination of the Lines of Constant Physics (LCPs). In order
to extract results valid in the continuum limit from numerical simulations on the lattice, the bare parameters of the action must be tuned
as the lattice spacing $a$ is reduced, so that renormalized physical observables remain invariant up to discretization effects. This is
typically achieved by fixing a set of hadronic observables. 

At finite temperature, however, when $T$ extends far beyond the hadronic scale $M_{had}$, such a strategy becomes impractical and
computationally prohibitive. In fact, the lattice must simultaneously accommodate both the hadronic and the thermal scale, requiring it to be fine
enough to resolve short distances while also large enough to capture long-range infrared dynamics. This leads to the hierarchy
$L \gg M_{had}^{-1} \gg T^{-1} \gg a$ which requires exceedingly large volumes in lattice units. This {\em window problem} effectively prevents the
use of usual hadronic renormalization schemes in the study of high-temperature thermodynamics. 

\begin{figure}[h]
  \centering
  \includegraphics[width=10cm]{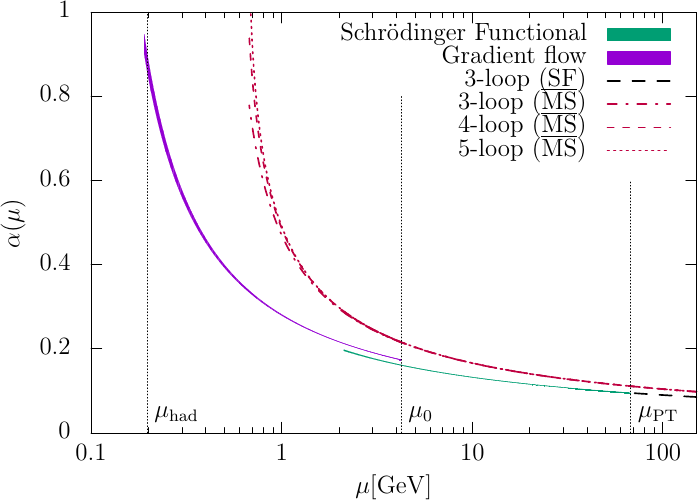}
  \caption{
    Non-perturbative running of renormalized gauge couplings for QCD with $N_f=3$ over a wide range of momentum scales. Plot taken from
    Ref.~\cite{Bruno:2017gxd}.   
  }
  \label{fig:alpha}
\end{figure}

To overcome this limitation, we adopt the strategy presented in Ref.~\cite{Giusti:2016iqr,DallaBrida:2021ddx}, in which the
LCPs are defined through a renormalized coupling computed non-perturbatively in a finite volume. The non-perturbative running of
renormalized couplings in the Schr\"odinger Functional (SF) and in the Gradient Flow schemes has been determined with high precision over a
wide range of energy scales~\cite{Luscher:1993gh,DallaBrida:2016uha,DallaBrida:2016kgh,DallaBrida:2018rfy}, as shown in
Fig.~\ref{fig:alpha} (taken from Ref.~\cite{Bruno:2017gxd}).
In particular, we consider the SF scheme~\cite{Luscher:1992an,Sint:1993un,Sint:1995ch}, where the renormalization scale $\mu$ is set by the
inverse box size, and the LCPs are defined by fixing the value of the renormalized coupling at a reference scale $\mu = 1/L_0$ for different
lattice spacings
\begin{equation}
    \bar{g}^2_{\rm SF}(g_0^2, a\mu) = \bar{g}^2_{\rm SF}(\mu)\,, \quad a\mu\ll 1\,.
\end{equation}
This construction enables a controlled evolution from the hadronic to the electroweak scale through a sequence of finite-volume
steps. In this framework, the bare parameters required for the high-temperature simulations are determined with high precision, as detailed in
Ref.~\cite{DallaBrida:2021ddx}.  

\subsection{Thermodynamics from a moving frame}
The standard approach to determining the EoS on the lattice is the {\em integral method}~\cite{Boyd:1995zg,Boyd:1996bx}. In
this framework, the pressure $p(T)$ is obtained by integrating the trace anomaly $\Theta(T)=e(T)-3p(T)$ over a range of temperatures
\begin{equation}
  \dfrac{p(T)}{T^4} = \dfrac{p(T_0)}{T_0^4} + \int_{T_0}^T \dfrac{\Theta (T')}{T'^5} dT'
  \label{eq:integral_method}
\end{equation}
where $e(T)$ is the energy density and $T_0$ is a reference temperature, typically chosen at low temperature where the pressure is
assumed to be negligible. Although this method has been highly successful in lattice thermodynamics, the trace anomaly, being a
dimension-four operator with a UV divergence $\propto a^{-4}$, requires subtracting a reference expectation value at the same lattice
spacing. If the reference and the target temperatures are significantly separated, the subtraction involves two different energy scales, which can make
the computation numerically demanding.  

This limitation can be circumvented by formulating QCD in a moving reference frame. As discussed in Eq.~(\ref{eq:shBCs}), this
corresponds to imposing shifted boundary conditions in the temporal direction with a non-vanishing spatial shift vector $\vxi$. The key advantage of
this geometric setup is that it enables a direct determination of the entropy density $s(T)$ without any temperature subtraction, so that
simulations are performed only at the temperature of interest.

The entropy density is obtained from the dependence of the free-energy density on the shift parameter, studying the thermodynamic response
of the system under variations of $\vxi$. By exploiting the relativistic invariance of the theory, it can 
be shown~\cite{Giusti:2012yj} that the thermodynamics of the shifted system is equivalent to that of a static frame at a temperature
determined by both the temporal extent $L_0$ and the shift vector $\vxi$. Defining the geometric factor $\gamma = 1/\sqrt{1+\vxi^2}$, the
inverse temperature is given by $T^{-1}=L_0/\gamma$. In the thermodynamic limit, the following relation holds between the free-energy
densities computed in the moving and in the static frames 
\begin{equation}
  f_\vxi(L_0) = -\lim_{L\rightarrow \infty} \frac{1}{L_0 L^3}\ln\cZ_\vxi (L_0,L,L,L) =
  -\lim_{L\rightarrow \infty} \frac{1}{L_0 L^3}\ln\cZ_{\vxi=0} (L_0/\gamma,L \gamma,L,L) =
  f_{\vxi=0}(L_0/\gamma).
\end{equation}
For simplicity, the shift vector is taken along one of the spatial directions. This identity reflects the underlying relativistic nature of
the shifted boundary conditions: introducing a non-vanishing shift is equivalent to performing a Lorentz boost in Euclidean space. The
geometric transformation has two simultaneous effects that the shift mixes: a time dilatation, where the effective temporal extent becomes
$L_0'=  L_0/\gamma$ and a spatial contraction in the direction of the shift $L'=  L \gamma$, such that the space-time volume element
$L_0 V = L_0' V' $ remains invariant. 

By keeping $L_0$ and $L$ fixed and varying the shift parameter, we can determine the entropy density through the expansion of the shift derivative
\begin{equation}
  \dfrac{\partial}{\partial \xi_k} = \dfrac {\partial (1/T)}{\partial \xi_k} \dfrac {\partial}{\partial (1/T)}
  +  \dfrac {\partial V'}{\partial \xi_k} \dfrac {\partial}{\partial V'}.
\end{equation}
Using the thermodynamic identity $T s(T) = e(T) + p(T)$, it follows that the derivative of the free energy density with respect to $\xi_k$
leads to the following fundamental relation with the entropy density
\begin{equation}\label{eq:slat}
  \frac{s}{T^3} = \frac{1+\vxi^2}{\xi_k}\frac{1}{T^4}\frac{\partial f_\vxi}{\partial \xi_k} \longrightarrow
  \frac{s}{T^3} = \frac{1+\vxi^2}{\xi_k}\frac{1}{T^4} \frac{\Delta f_\vxi}{\Delta\xi_k}.
\end{equation}
The expression on the right represents the lattice discretization of the derivative, which allows a non-perturbative computation of
the entropy density through numerical simulations.

A fully non-perturbative determination of the entropy density can also be obtained from the expectation values of matrix
elements of the renormalized energy-momentum tensor $T_{\mu\nu}$ on the lattice. In the lattice regularization, the explicit breaking of
translational and rotational invariance implies that $T_{\mu\nu}$ is no longer protected by the usual continuum conservation laws and finite
renormalization constants are required to recover the correct properties in the continuum limit. At present, these constants are known 
only at one-loop order in perturbation theory~\cite{Caracciolo:1991vc,Caracciolo:1991cp} and a fully non-perturbative determination is still lacking. 

\section{Numerical computation}\label{sec:Numerical computation}
The entropy density is computed using Eq.~(\ref{eq:slat}) employing the two-point symmetric discretization for the derivative of the
free-energy density with respect to the $k$-th component of the shift
\begin{equation}
  \frac{\Delta f_\vxi}{\Delta\xi_k} = 
  \frac{L_0}{4a}\left(f_{\vxi+\frac{2a}{L_0}\hat{k}} - f_{\vxi-\frac{2a}{L_0}\hat{k}}\right).
  \label{eq:f_diff_discrete_shift}
\end{equation}
From a computational perspective, it is advantageous to decompose the discrete derivative of the free-energy density into two separate
contributions at fixed bare parameters
\begin{equation}
  \frac{\Delta f_\vxi}{\Delta\xi_k} = 
  \frac{\Delta f_\vxi^\infty}{\Delta\xi_k} + 
  \frac{\Delta (f_\vxi - f_\vxi^\infty)}{\Delta\xi_k}\,,
  \label{eq:split}
\end{equation}
where $f_\vxi^\infty$ denotes the free-energy density in the limit of infinitely heavy quarks, corresponding to the SU(3) pure gauge theory.
The first term on the right-hand side is evaluated by rewriting it as an integral over the bare coupling $g_0^2$
\begin{equation}
  \frac{\Delta f_\vxi^\infty}{\Delta\xi_k} = 
  \frac{\Delta f^{(0),\infty}}{\Delta\xi_k}
  + g_0^2\, \frac{\Delta f^{(1),\infty}}{\Delta\xi_k} 
  - \int_0^{g_0^2} du \left(\frac{1}{u}\left.\frac{\Delta\corrno{\overline{S_G}}_\vxi^\infty}{\Delta\xi_k}\right|_{g_0^2=u}
    +\frac{\Delta f^{(1),\infty}}{\Delta\xi_k}\right)\,.
  \label{eq:Df_gauge}
\end{equation}
where $f^{(0),\infty}$ and $f^{(1),\infty}$ denote, respectively, the tree-level and the one-loop coefficients of the expansion in
lattice perturbation theory of the pure gauge free-energy density~\cite{DallaBrida:2020gux}. The integral involves the expectation value of
the SU(3) pure gauge action density in the moving reference frame 
\begin{equation}
  \corr{\overline{S_G}}^{\infty}_\vxi = \frac{a^4}{L_0L^3}\corr{S_G}^{\infty}_\vxi.
  \label{eq:corrSG_volavg}
\end{equation}
and the subtraction of the one-loop coefficient has the purpose of ensuring a numerically more stable integration from the weak-coupling limit. 

The second term in Eq.~\eqref{eq:split} accounts for the dynamic contribution of the three massless quark flavours. At fixed system size and
bare gauge coupling, it can be determined by
integrating the expectation value of the scalar density $\corrno{\psibar\psi}_\vxi^{m_q}$ -- which represents the response of the system to
variations in the quark mass -- over a range connecting the chiral and the static limits
\begin{equation}
   \frac{\Delta (f_\vxi - f_\vxi^\infty)}{\Delta\xi_k} =  
   - \frac{\Delta}{\Delta\xi_k} \int_0^\infty dm_q\frac{\partial f_\vxi^{m_q}}{\partial m_q} 
   = -\int_0^\infty dm_q \frac{\Delta\corrno{\psibar\psi}_\vxi^{m_q}}{\Delta\xi_k}\,,
   \label{eq:Df_quark}
\end{equation}
where $m_q = m_0 - m_{\rm cr}(g_0^2, L_0)$ and $m_{\rm cr}$ is the critical mass. 

The numerical evaluation of the entropy density thus relies on the lattice determination of the two above contributions and Monte Carlo
simulations have been performed at the set of temperatures reported in Table~\ref{tab:T0T8GeV}, corresponding to fixed values of the
Schr\"odinger Functional coupling at the scale $\mu=1/L_0$. The shift vector entering the definition of the entropy density has
been set to $\vxi=(1,0,0)$. For each temperature, the spatial size has been fixed to $L/a=144$ while
four lattice resolutions in the temporal direction have been considered, $L_0/a=4,6,8,10$, corresponding to values for $(a T)^{-1}$ between
5.6 and 14. This choice ensures aspect ratios $LT$ in the range
$\sim 10$--$25$, sufficiently large to suppress finite-volume effects.

\subsection{Determination of \mtht{$\static$}}
\begin{figure}[h]
    \centering
    \includegraphics[width=10cm]{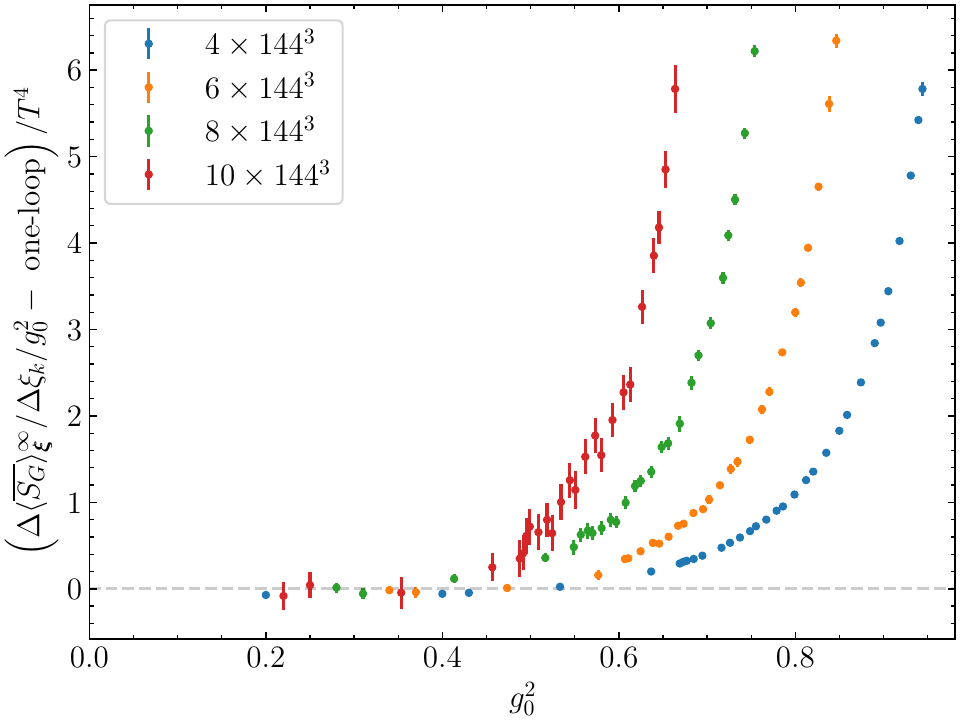}
    \caption{
    Plot of the integrand function in Eq.~\eqref{eq:Df_gauge} as a function of the bare coupling $g_0^2$.
    Points have been shifted horizontally by $0.03\times(L_0/a-4)$ for better readability.
    }
    \label{fig:g02_integrand_06x144}
\end{figure}

The contribution of ${\Delta f_\vxi^\infty}/{\Delta\xi_k}$ is obtained computing the integral over the bare coupling appearing in
Eq.~\eqref{eq:Df_gauge}. The corresponding 
integrand is a smooth function of $g_0^2$, allowing for an accurate numerical evaluation through a suitable combination of quadrature
rules. The integration domain in $g_0^2$ is partitioned into intervals where different quadrature schemes are applied, including Simpson,
trapezoidal and Gauss--Legendre rules. At fixed $L_0/a$, the computation proceeds recursively in the temperature: once the integral is
determined up to a given value of the bare coupling, the result at lower temperatures is obtained by adding the contribution from the
subsequent interval in 
$g_0^2$. This strategy minimizes the number of independent integrations while maintaining high numerical accuracy. 
At each quadrature point, the expectation value of the gauge action density is computed in pure SU(3) Yang--Mills theory at the two values
of the shift given in Eq.~(\ref{eq:f_diff_discrete_shift}). The specifics of the integration scheme are detailed in Table~\ref{tab:tab_g02_integral}.
\begin{table}[b]
    \centering
    \begin{tabular}{|c|lc|}
        \hline
        Interval & \multicolumn{2}{c|}{Quadrature} \\
        \hline
        \multirow{2}{*}{$0\leq g_0^2\leq 6/15$} & 3 (Simpson) & $L_0/a=4$ \\
                                                & 2 (trapezoid) & $L_0/a=6,8,10$ \\
        \hline
        $6/15 \leq g_0^2 \leq 6/9$              & 3 (Gauss-Legendre) &    \\
        \hline
        \multirow{2}{*}{$6/9\leq g_0^2\leq g_0^2|_{T_0}$} & 3 (Gauss-Legendre)    & $L_0/a=4$ \\
                                                          & 1 (midpoint) & $L_0/a=6$ \\
        \hline
        $6/9 \leq g_0^2 \leq g_0^2|_{T_1}$                & 3 (Gauss-Legendre)  & \\
        \hline
        \multirow{2}{*}{$g_0^2|_{T_{i-1}}\leq g_0^2\leq g_0^2|_{T_i}$} & 3 (Gauss-Legendre)    & $1<i<7$ \\
                                                                      & 5 (Gauss-Legendre)    & $i=7,8$ \\
        \hline
    \end{tabular}
    \caption{Summary of the integration scheme for the computation of the integral in $g_0^2$ appearing in Eq.~\eqref{eq:Df_gauge}.}
    \label{tab:tab_g02_integral}
\end{table}
Gauge ensembles are generated using a combination of heat-bath and over-relaxation updates implemented via the Cabibbo--Marinari procedure. 
The statistical precision of this contribution depends on the lattice resolution. Coarser lattices achieve permille-level accuracy, while at
the finest lattice spacing the uncertainty increases to the percent level, reflecting the higher computational cost. The integrand function
of Eq.~\eqref{eq:Df_gauge} is illustrated in Figure~\ref{fig:g02_integrand_06x144} for the four lattice resolutions $L_0/a=4,6,8,10$. The final
results can be found in Table II in Ref.~\cite{Bresciani:2025vxw}.

\subsection{Determination of \mtht{$\dynamic$}}
The contribution of ${\Delta (f_\vxi - f_\vxi^\infty)}/{\Delta\xi_k}$ is evaluated from the integral over the bare subtracted quark mass
in Eq.~\eqref{eq:Df_quark}. The integration 
range is split into three regions in terms of the dimensionless variable $\mt_q=m_q/T$, allowing for an optimized representation of the integrand
over the full mass range from the chiral to the static limit. Gaussian quadratures of different orders are employed in each region, ensuring
that systematic effects from the numerical integration remain negligible compared to the statistical uncertainties. In the large-mass region, a
change of variables to the hopping parameter makes the integration domain finite and improves numerical stability. The computation requires
evaluating the scalar density expectation value at a set of quark masses corresponding to the quadrature nodes. For each lattice setup,
simulations are performed at multiple mass values and for two shift vectors, resulting in a sizable number of independent ensembles.

A key optimization arises from the behaviour of the fermionic sector at large quark masses. As the quark mass increases, fermionic
contributions become progressively suppressed, allowing a coarser integration of the molecular dynamics in the Hybrid Monte Carlo
algorithm without degrading acceptance rates. This feature is exploited to significantly reduce computational costs. Moreover, a variance
reduction in the heavy quark mass regime is achieved through introducing an improved estimator~\cite{Giusti:2019kff} of the chiral 
condensate obtained by subtracting the leading non-trivial order in the hopping parameter expansion. 

\begin{figure*}[t]
  \begin{center}
    \includegraphics[width=7.5cm]{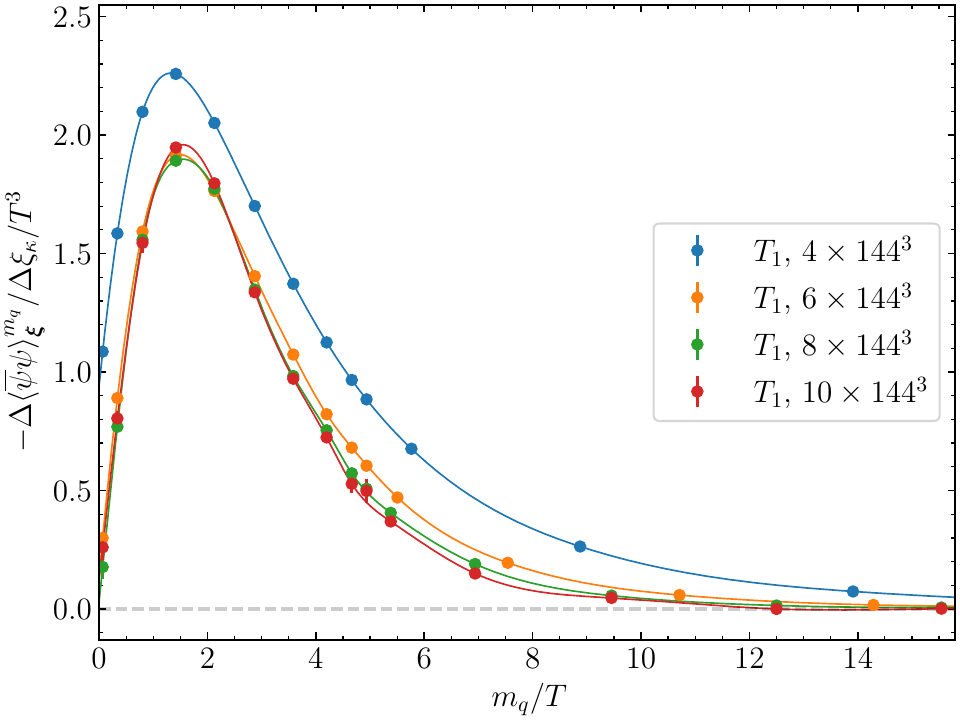}
    \includegraphics[width=7.5cm]{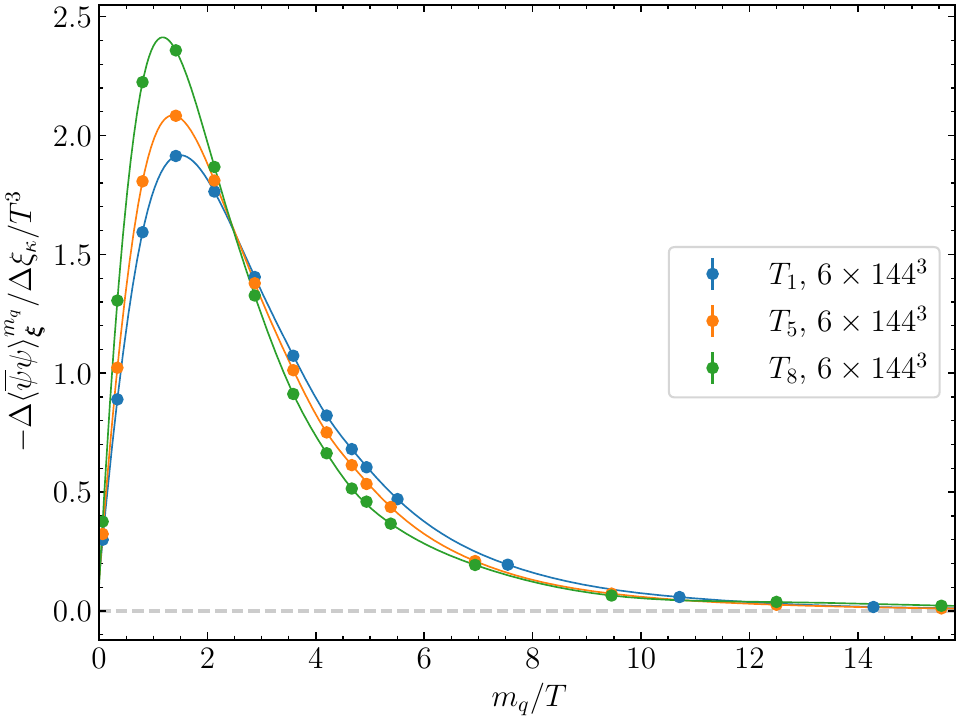}
  \end{center}
  \caption{
    Left: plot of the integrand function in Eq.~\eqref{eq:Df_quark}
    computed at the bare parameters of temperature $T_1$ and at the resolutions $L_0/a=4,6,8,10$, 
    as a function of $m_q/T$. Points have been interpolated with a cubic spline to guide the eye.
    In most cases, errors are smaller than the markers.
    Right: the same integrand function is shown at the resolution $L_0/a=6$ and at three temperatures.
  }
  \label{fig:mq_integrand_06x144}
\end{figure*}

Finally, the computational effort at the various quadrature points is optimized by taking into account both the variance of the
observable and the weight of each node in the integration formula. This strategy ensures that the overall uncertainty is minimized
for a fixed computational budget, yielding a significant gain in efficiency compared to collecting uniform statistics across all mass values. 
The non-perturbative integrand function is represented in Figure~\ref{fig:mq_integrand_06x144} for different values of the lattice spacing
(left panel) and at different temperatures (right panel).
Overall, the combination of optimized integration schemes, algorithmic improvements, and variance reduction techniques enables a precise
determination of both contributions to the entropy density over the full range of temperatures and lattice spacings considered. The final
results can be found in Table II in Ref.~\cite{Bresciani:2025vxw}.

\subsection{Extrapolation to the continuum limit}
\begin{figure}[t]
    \centering
    \includegraphics[width=10cm]{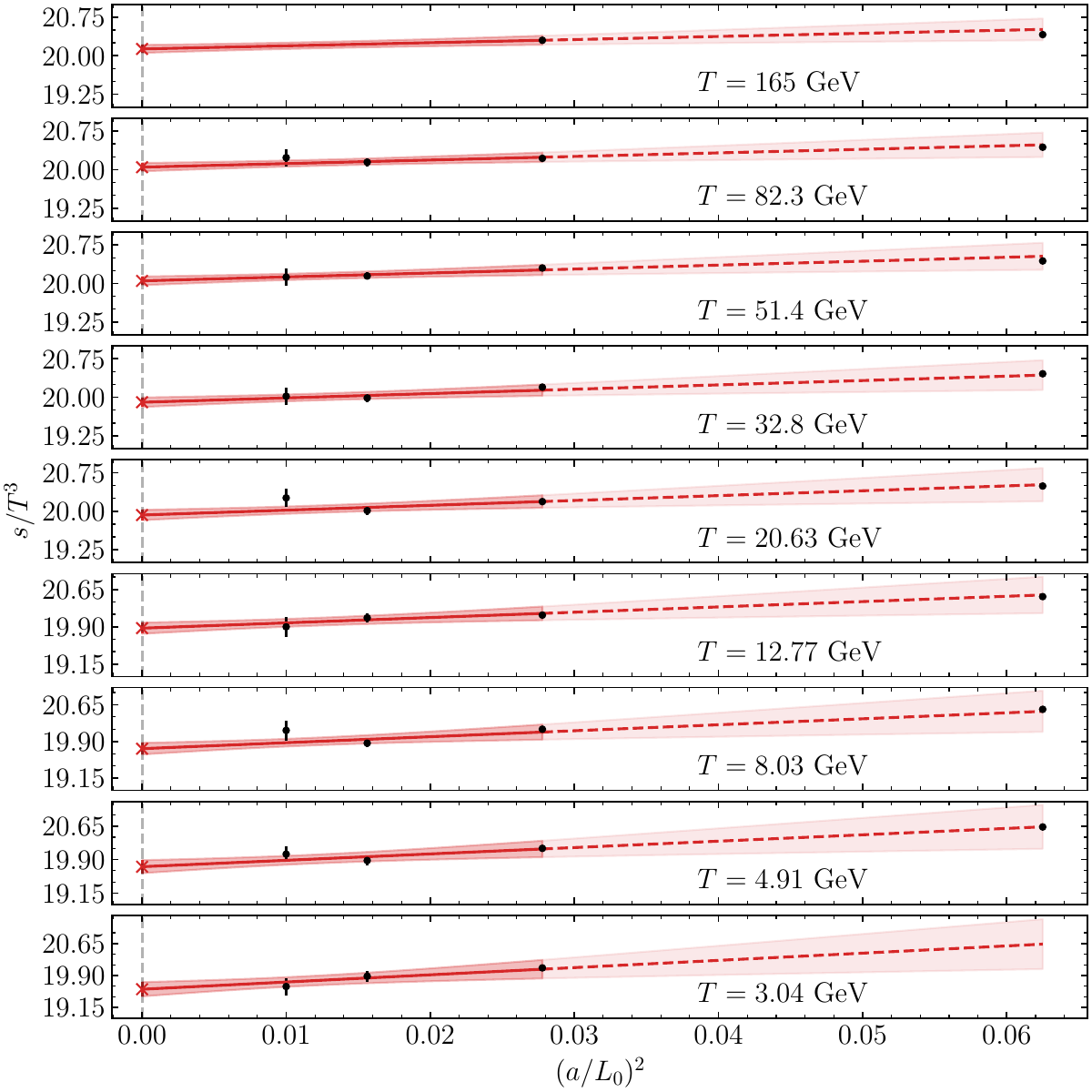}
    \caption{Black dots are the values of the one-loop improved entropy density as a function of $(a/L_0)^2$ 
    at the temperatures $T_0, T_1, ..., T_8$. The red band is our best extrapolation to the continuum limit and red crosses are the
    continuum extrapolated values for $s/T^3$. The horizontal axis is common to all the subplots.
    }
    \label{fig:clim}
\end{figure}

The final results for the discrete derivative of the free energy density with respect to the shift can be found in Table III in
Ref.~\cite{Bresciani:2025mcu} and the continuum limit is obtained after implementing a one-loop perturbative improvement that removes
lattice artifacts at tree-level and at O($g^2$). The remaining cutoff effects are therefore suppressed and
expected to start at O($a^2 g^3$), consistently with O($a$)-improvement and the presence of odd powers in the coupling at finite
temperature. The extrapolation is performed with a global fit of the data at all temperatures, in which discretization effects are modeled as
polynomials in $\bar g_{SF}$. A careful analysis of fit stability shows that the coarsest lattice
$L_0/a=4$ is significantly affected by higher-order discretization artifacts. It is therefore excluded from the preferred fit but its impact
is taken into account to estimate the systematic uncertainty associated to higher-order corrections in the lattice spacing.
The final estimate uses data at $L_0/a=$ 6, 8 and 10 and a minimal ansatz for cutoff effects, yielding
continuum values with sub-percent precision. Several variations of the fit ansatz, including higher-order terms in $a$ and in the coupling,
as well as logarithmic corrections, lead to results fully compatible within uncertainties, showing that the continuum extrapolation is
robust and that residual discretization effects are well under control.
The final continuum results for the entropy density are reported in Table~\ref{tab:T0T8GeV} and have a relative error of
$0.5$-$1.0\%$ and Figure~\ref{fig:clim} displays the related continuum limit~extrapolations. 
\begin{table}
  \centering
  \begin{tabular}{|c|c|c|c|}
    \hline
    $T$  & $T$ (GeV) &  $\bar{g}^2_{\rm SF}(\mu=1/L_0)$ & $s/T^3$\\
    \hline
    $T_0$ &  164.6(5.6) &  1.01636 &  20.13(8)  \\
    $T_1$ &  82.3(2.8)  &  1.11000 &  20.05(8)  \\
    $T_2$ &  51.4(1.7)  &  1.18446 &  20.05(9)  \\
    $T_3$ &  32.8(1.0)  &  1.26569 &  19.90(9)  \\
    $T_4$ &  20.63(63)  &  1.3627  &  19.93(10) \\   
    $T_5$ &  12.77(37)  &  1.4808  &  19.87(11) \\
    $T_6$ &  8.03(22)   &  1.6173  &  19.75(12) \\
    $T_7$ &  4.91(13)   &  1.7943  &  19.74(15) \\
    $T_8$ &  3.040(78)  &  2.0120  &  19.58(17) \\
    \hline
  \end{tabular}
  \caption{Second column: physical temperatures considered in this study. 
    Third column: values of the Schr\"odinger functional coupling in $\Nf=3$ QCD at the renormalization scale $\mu=1/L_0$. Fourth column:
    values of the entropy density in the continuum limit.}
  \label{tab:T0T8GeV}   
\end{table}

Finite-volume effects in high-temperature lattice simulations are exponentially suppressed as $e^{-M_{\rm gap} L}$, where $M_{\rm gap}$ 
is the mass of the lightest screening state. In the temperature regime investigated here, the screening mass is approximately of the order
of the physical temperature~$T$. Given that the simulations were performed on lattices with aspect ratios ranging between $10\lesssim
LT\lesssim 25$, finite-size corrections are expected to be negligible compared to the statistical precision of our results. This expectation
was explicitly validated through dedicated simulations at varying spatial volumes, which showed no statistically significant deviations in
the thermodynamic observables.

Concerning the topological sampling, the dominance of the trivial sector at high temperatures justifies restricting the measurements
accordingly. While occasional tunneling between topological sectors was observed at the lowest temperatures and coarsest lattice spacings in
the pure gauge simulations, no statistically significant dependence of the action density on the topological charge was
detected. Nevertheless, to account for potential residual effects in the expectation value of the action density due to limited
topological sampling, a conservative systematic uncertainty of up to 2\% has been included in the error estimates of
$\Delta\corrno{\overline{S_G}}_\vxi^\infty$ for those cases.

\section{Equation of State}
\label{sec:Equation of State}
The data generated by the Monte Carlo simulations and the extrapolation of the results to the continuum limit allowed to investigate the
temperature dependence of the thermodynamic potentials over the broad range of temperatures between 3~GeV and
165~GeV~\cite{Bresciani:2025vxw,Bresciani:2025mcu}. For convenience, we express the results in terms of $\hat g(\mu)$ defined as the
five-loop $\MSbar$ coupling~\cite{Baikov:2016tgj} at the renormalization scale $\mu=2\pi T$, whose leading order expression is  
\begin{equation}\label{eq:gmu}
  \frac{1}{\hat g^2(\mu)} = \frac{9}{8\pi^2} \ln  \frac{\mu}{\Lambda_{\MSbar}} +\ldots \, , 
  \qquad \mu=2\pi T\,,
\end{equation}
where $\Lambda_{\MSbar} = 341$~MeV is taken from Ref.~\cite{Bruno:2017gxd}. For our purposes, this is only a convenient function of $T$ for
studying the temperature dependence of the non-perturbative results and which, at the same time, simplifies the comparison with the 
perturbative expectations. 

\begin{figure}[ht]
    \centering
    \includegraphics[width=10cm]{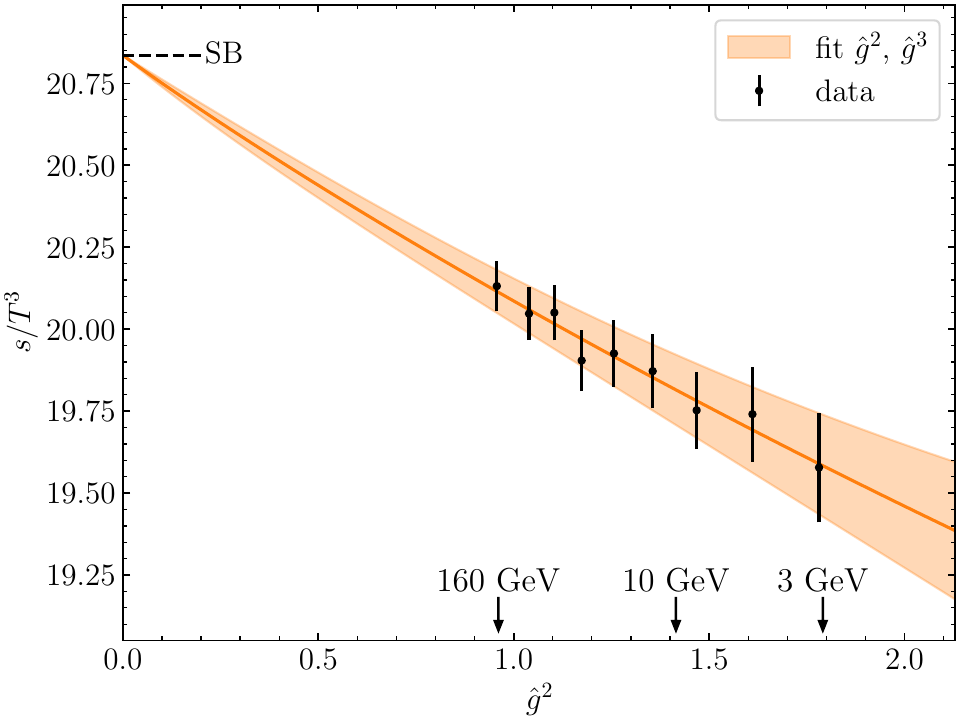}
    \caption{Temperature dependence of the entropy density in terms of $\hat g^2$. The data are fitted by enforcing the Stefan--Boltzmann
      limit (SB) at infinite temperature and including quadratic and cubic corrections. The shaded area indicates the uncertainty of the fit.} 
    \label{fig:bests_comp}
\end{figure}

The continuum-extrapolated results for the entropy density have a remarkably smooth dependence on $\hat g^2$ suggesting that it may be
an effective variable for characterizing the thermal behaviour of the theory in the high-temperature regime. In order to describe the data
and to facilitate the comparison with perturbation theory, we parametrize the entropy density using a polynomial expansion 
\begin{equation}
  \frac{s}{T^3} = \frac{32\pi^2}{45}\sum_k s_k\left(\frac{\hat{g}}{2\pi}\right)^k\,.
  \label{eq:s_mikko}
\end{equation}
Several fit strategies have been explored in order to assess systematic uncertainties and the sensitivity to the asymptotic 
behaviour at high temperature. We can enforce the Stefan--Boltzmann limit (SB) at asymptotically large $T$
fixing the leading coefficient to the ideal gas value $s_0=s_0^{\rm SB}=2.969$ (for $N_f=3$) and leaving the first two non-trivial
coefficients as fit parameters. As shown in Fig.~\ref{fig:bests_comp}, this parametrization provides an excellent description of the data
across the investigated temperature range and the estimated values of the parameters are $s_2=-5.1(9)$ and $s_3=5(5)$. 
While the quality of the fit is high, a first comparison with perturbation theory shows a significant deviation with respect to the expected
value of the quadratic coefficient $s_2=-8.438$.

\begin{figure}[ht]
\begin{center}
\includegraphics[width=7.5cm]{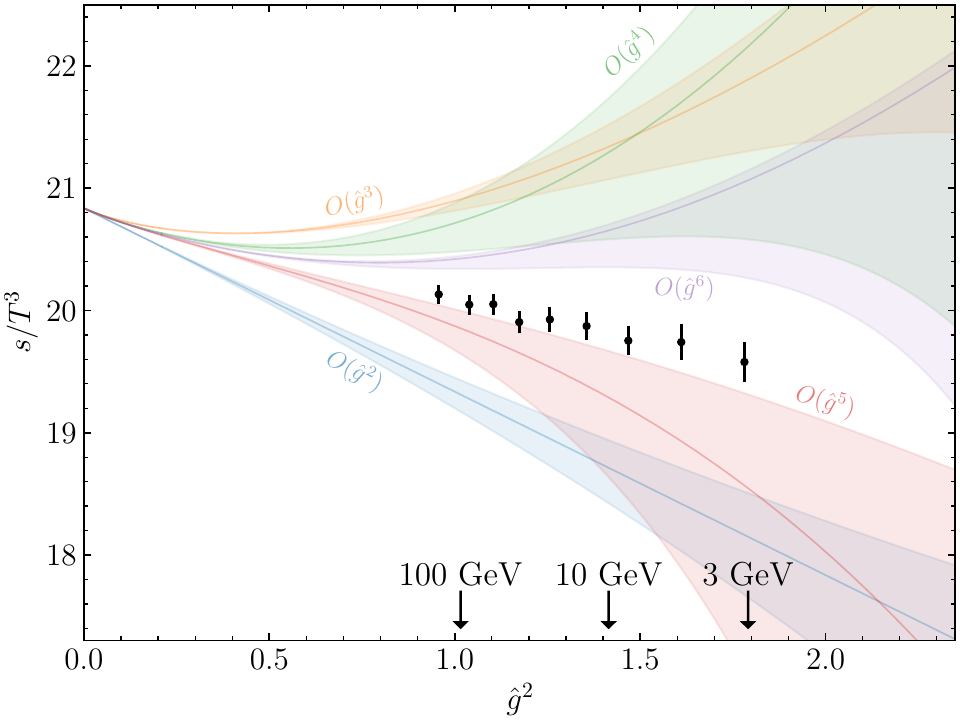}
\includegraphics[width=7.5cm]{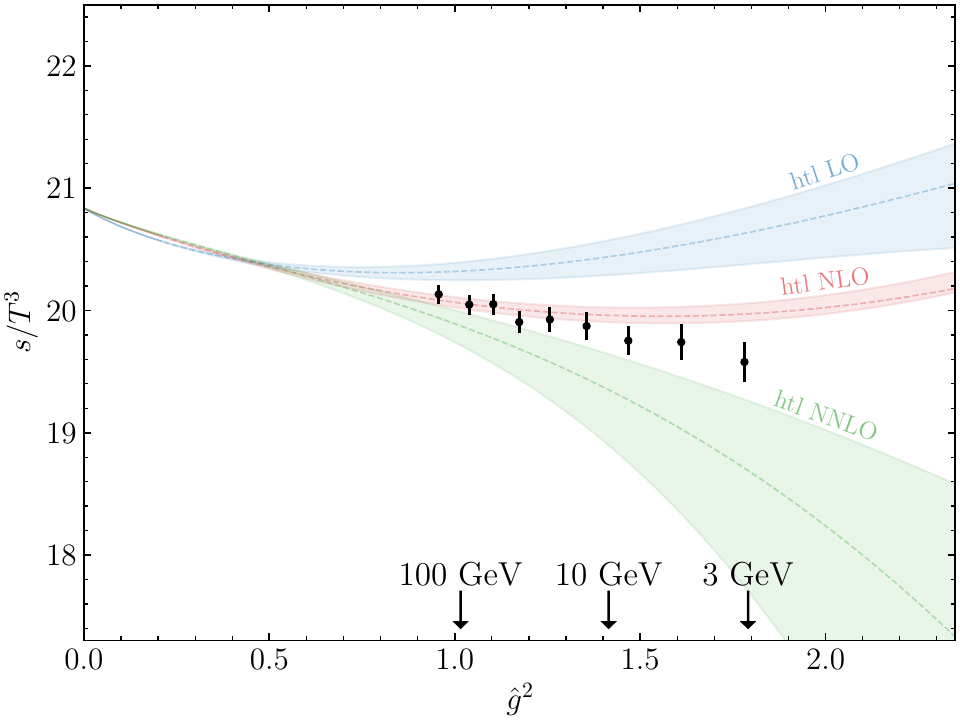}
\caption{Black dots are the non-perturbative values of $s/T^3$ plotted against $\hat g^2$.
Some values of the physical temperature are reported for reference.
In the left plot, the curves represent the predictions of perturbation theory
obtained from Ref.~\cite{Kajantie:2002wa}, each including up to the order in $\hat g$ indicated by the label.
In the right plot, the curves represent the hard thermal loop perturbation 
theory~\cite{Andersen:2003zk,Andersen:2010ct,Andersen:2011sf}
at leading order (LO), next-to leading order (NLO) and next-to-next-to leading order (NNLO).
In both plots, error bands are obtained by a variation of the renormalization scale $\mu=2\pi T$ by the factors $0.5$ and $2$.}
\label{fig:comp_pt_htl}
\end{center}
\end{figure}

We have then compared -- see Fig.~\ref{fig:comp_pt_htl} -- the non-perturbative results for $s/T^3$ with the perturbative predictions of
high-temperature QCD, computed both in standard perturbation theory and in hard thermal loop (HTL) resummed approaches, respectively known
up to O$(g^6 \ln g^2)$ and to NNLO.  
In conventional perturbation theory, the expansion exhibits poor convergence in the considered temperature range. Successive orders 
show significant deviations, indicating that higher-order contributions are relevant even at temperatures as high as
$T \sim 100$~GeV. HTL-resummed calculations improve the apparent convergence, especially at the highest temperatures, but still display visible 
tensions between successive orders. The comparison with the non-perturbative data shows that even the most accurate HTL results 
do not fully capture the dynamics in the explored regime. 

\begin{figure}[ht]
\begin{center}
\includegraphics[width=7.5cm]{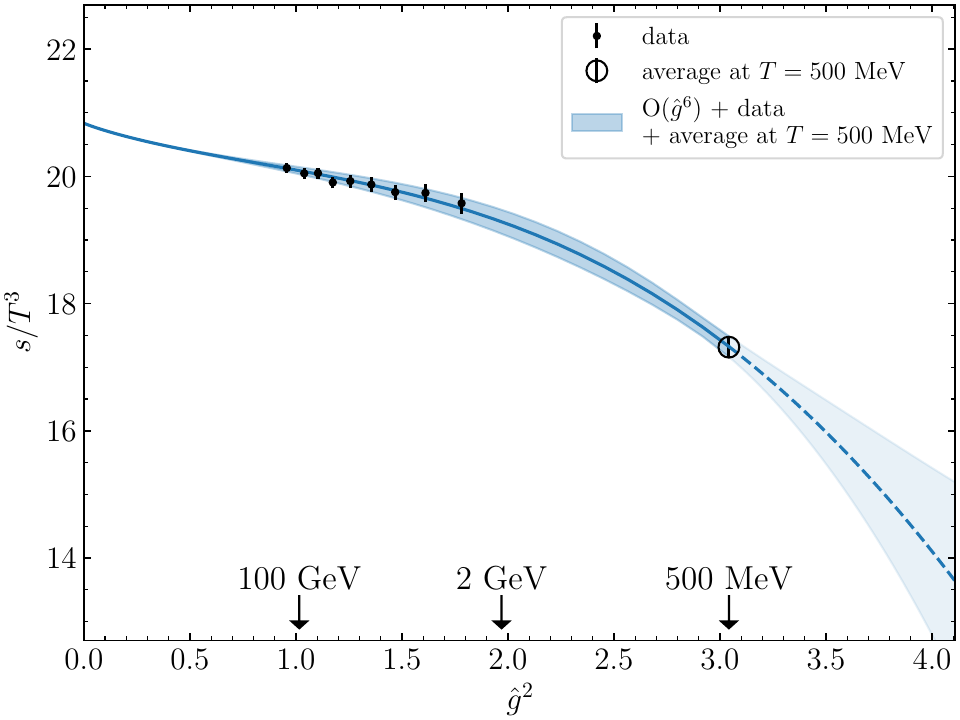}
\includegraphics[width=7.5cm]{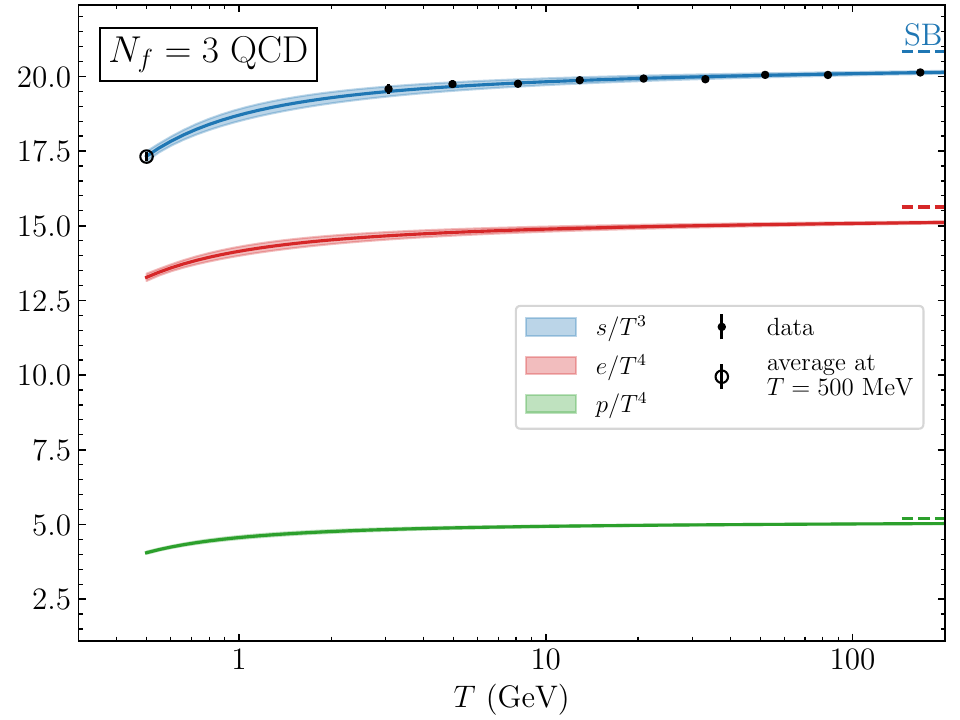}
\caption{Left: entropy density $s/T^3$ in the continuum limit as a function of $\hat g^2$.
         The blue band is the best parametrization for $T\geq500$ MeV, and includes the point at $T=500$ MeV obtained 
         from Refs.~\cite{Borsanyi:2013bia,HotQCD:2014kol,Bazavov:2017dsy} (open circle marker).
       Right: EoS as a function of the temperature for $T\geq 500$ MeV.}
\label{fig:best_sep}
\end{center}
\end{figure}

As alternative fit strategy we have incorporated all known perturbative coefficients, fitting only the unknown higher-order
terms. It is important to note that at O$(g^6)$ non-perturbative contributions to the pressure appear that cannot be computed in the framework of
perturbation theory~\cite{Kajantie:2002wa}. Our preferred parametrization of the non-perturbative data involves $\hat g^6$ and $\hat g^7$
terms and includes the entropy density at $T=500~\mathrm{MeV}$ obtained from independent lattice calculations with $N_f=2+1$
flavours~\cite{Borsanyi:2013bia,HotQCD:2014kol,Bazavov:2017dsy}. This input effectively constraints the interpolation at lower temperatures and
leads to a global description of the entropy density valid for $T \geq 500$~MeV. This phenomenological fit is shown in the left panel of
Fig.~\ref{fig:best_sep}. 

Using this parametrization of the entropy density, the pressure and energy density can be obtained consistently. 
Analogously to Eq.~(\ref{eq:s_mikko}), we parametrize the pressure as a polynomial expansion and its coefficients are directly related to
those of the entropy density through the thermodynamic identity $s(T) = \frac{d}{dT}p(T)$. The energy density
is then obtained from $e(T) = T s(T) - p(T)$. As a cross-check, we have computed the pressure by integrating the entropy density starting
from $T_0 = 500$~MeV and $p(T_0)/T_0^4=4.050(38)$ taken from independent lattice determinations~\cite{Borsanyi:2013bia,HotQCD:2014kol,Bazavov:2017dsy}. 
The two procedures yield fully compatible results across the entire temperature range, providing a non-trivial validation of the consistency
of the analysis. The temperature dependence of the thermodynamic potentials is shown in the right panel of Fig.~\ref{fig:best_sep}.

\section{Conclusions and Outlook}

The results presented in these proceedings show that current theoretical and computational advances in lattice QCD permit a fully
non-perturbative investigation of QCD thermodynamics up to the electroweak scale. The QCD Equation of State for $N_f=3$ massless quarks has
been computed with a precision of 0.5-1\% across the previously unexplored temperature range between 3 GeV and 165
GeV~\cite{Bresciani:2025vxw,Bresciani:2025mcu}.  

An important issue of this study is the behaviour of the entropy density as it approaches the Stefan--Boltzmann limit. While $s/T^3$
exhibits a relatively mild and smooth decrease from the ideal gas limit down to temperatures of a few GeV, this regularity should not be
(mis)taken for an indication that the system is in a perturbative regime. In fact, our analysis shows that terms beyond the known
O$(g^6)$ perturbative expansion that include non-perturbative contributions are necessary to describe the lattice data accurately up to the
electroweak scale. Interestingly, we find that a phenomenological fit based on these extensions remains robust even down to $T\sim 500$~MeV. 

The strategy developed for this study, combining shifted boundary conditions with the non-perturbative running of the Schr\"odinger
Functional coupling, represents an efficient approach that can be applied to include the cases of $N_f=4$ and 5 massive flavours without
further theoretical development, providing a clear path toward a complete description of the Standard Model's thermodynamic evolution in the
early Universe. 

Looking ahead, simulations are currently in progress to determine the non-perturbative renormalization constants of the energy-momentum
tensor. They are essential for the calculation of transport coefficients whose accurate determination will be important for understanding the 
hydrodynamical features of the quark-gluon plasma in both heavy-ion collisions and cosmological contexts.

\vspace{.5cm}
\noindent
{\bf Acknowledgements.} The results reported here were obtained with M.~Bresciani, M.~Dalla~Brida, and L.~Giusti, whom I gratefully
acknowledge for their collaboration and for a critical reading of these proceedings. I also thank CINECA for providing us with a very
generous access to Leonardo during the early phases of operations of the machine and for the computer time allocated via the CINECA-INFN,
CINECA-Bicocca agreements. The R\&D has been carried out on the PC clusters Wilson and Knuth at Milano-Bicocca. We thank all these
institutions for the technical support. This work is (partially) supported by ICSC – Centro Nazionale di Ricerca in High Performance
Computing, Big Data and Quantum Computing, funded by European Union – NextGenerationEU.  

\bibliographystyle{JHEP}
\bibliography{bibfile}

\end{document}